\documentclass[twocolumn,prb,aps,showpacs]{revtex4-1}
\usepackage{amsmath}
\usepackage{mathrsfs}
\usepackage{amssymb}
\usepackage{graphics}
\usepackage{dcolumn}
\usepackage{bm}
\usepackage{graphicx}
\usepackage{dcolumn}
\usepackage{bm}
\usepackage{hyperref}
\selectfont

\newcommand{\beq}{\begin{equation}}
\newcommand{\eeq}{\end{equation}}
\newcommand{\beqa}{\begin{eqnarray}}
\newcommand{\eeqa}{\end{eqnarray}}
\usepackage{graphicx}
\usepackage{amsmath}
\usepackage{bm}

\begin{document}

\title{Goos-H\"{a}nchen-like shifts for Dirac fermions in monolayer graphene barrier}

\author{Xi Chen$^{1,2}$}

\email{xchen@shu.edu.cn}

\author{Jia-Wei Tao$^{1}$}

\author{Yue Ban$^{2}$}


\affiliation{$^{1}$ Department of Physics, Shanghai University,
200444 Shanghai, China}

\affiliation{$^{2}$ Departamento de Qu\'{\i}mica-F\'{\i}sica,
UPV-EHU, Apdo 644, 48080 Bilbao, Spain}

\date{\today}

\begin{abstract}
We investigate the Goos-H\"{a}nchen-like shifts for Dirac fermions in transmission through a monolayer graphene barrier. The lateral shifts, as the functions of the  barrier's width and the incidence angle, can be negative and positive in Klein tunneling and classical motion, respectively. Due to their relations to the transmission gap, the lateral shifts can be enhanced by the transmission resonances when the incidence angle is less than the critical angle for total reflection, while their magnitudes become only the order of Fermi wavelength when the incidence angle is larger than the critical angle. These tunable beam shifts can also be modulated by the height of potential barrier and the induced gap, which gives rise to the applications in graphene-based devices.

\end{abstract}



\maketitle


\section{Introduction}

Monolayer graphene has attracted much attention
\cite{Neto-GPN,Beenakker} since the graphitic sheet of one-atom
thickness has been experimentally realized by A. K. Geim \textit{et
al.} in 2004 \cite{Novoselov-GMJ}. The valence electron dynamics in
such a truly two-dimensional (2D) material is governed by a massless
Dirac equation. Thus graphene has many unique electronic and
transport properties \cite{Neto-GPN,Beenakker}, including Klein tunneling
\cite{Katsnelson-NG}. Recently, further investigations show that a method to produce
a finite bandgap in graphene sheets by epitaxially on proper substrate \cite{Zhou} has been proposed,
and the Dirac fermions in gapped graphene are described by 2D massive Dirac equations.
The induced gap at the Dirac point is significant to control the transport of the carriers
and integrating graphene into the semiconductor technology.

Motivated by these
progress, the transport of massive or massless Dirac fermions in graphene
opens a way to design various graphene-based electron devices in term of the electron optics
behaviors \cite{Cheianov,Park,Darancet,Ghosh,Zhao,Beenakker-PRL}, such as
focusing \cite{Cheianov}, collimation \cite{Park}, Bragg reflection \cite{Ghosh}, and Goos-H\"{a}nchen effect (GH)
\cite{Zhao,Beenakker-PRL}. In particular, Zhao and
Yelin \cite{Zhao} have shown that, based on the electronic
counterpart to the trapped rainbow effect in optics
\cite{Tsakmakidis}, the interplay GH effect and negative refraction
\cite{Cheianov} in graphene leads to the coherent graphene devices,
such as movable mirrors, buffers and memories. Beenakker \textit{et
al.} \cite{Beenakker-PRL} have further found that the GH effect at a
$n$-$p$ interface in graphene doubles the degeneracy of the lowest
propagating mode, which can be observed as a stepwise increase by
$8e^2/h$ of the conductance with increasing channel width.

In this paper, we will investigate the negative and positive
lateral shifts for Dirac fermions in transmission through a 2D monolayer graphene
barrier, based on the tunable transmission gap \cite{XChen-APL}. Generally, the magnitude of
GH shift for total reflection in graphene is in the order of Fermi wavelength.
However, the lateral shifts discussed here are similar to but different from the
conventional GH shift, because they do result from
the beam width of Dirac fermions, and can be enhanced by the transmission resonances.
As a matter of fact, the lateral shifts in transmission have nothing to do with the evanescent wave,
thus we term them as Goos-H\"{a}nchen-like (GHL) shift, which can be considered as an electronic
analog of the lateral shifts in the optics \cite{XChen-PRE} and atom optics \cite{Zhang-PRA}.
More interestingly, not only large positive but also large negative GHL shifts can occur in
transmitted beam through the graphene barrier.
The negative shift behaves like the phenomena of negative refraction in graphene  \cite{Cheianov}.
In addition, we will also discuss the effect of the induced gap on the GHL shifts in the gapped graphene barrier.
All these tunable beam shifts in the monolayer graphene barrier can be
applied in the design of graphene-based devices, such as electron wave switch, wave vector or energy filters and splitter.

\begin{figure}[]
\scalebox{0.8}[0.8]{\includegraphics{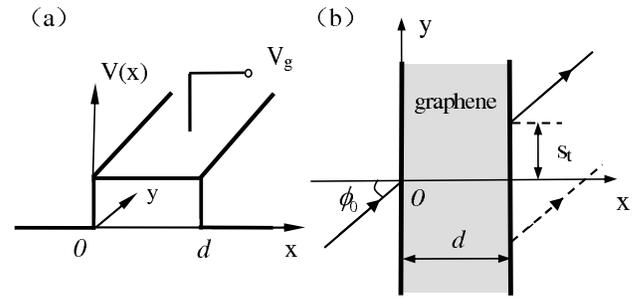}}\caption{ (a) Schematic
diagram for two-dimensional monolayer graphene barrier. (b) Negative and positive lateral shifts of Dirac fermions in transmission. \label{fig.1}}
\end{figure}

\section{Theoretical Model}
Consider the massless Dirac fermions with Fermi energy $E$ at angle
$\phi_0$ with respective to the $x$ axis incident from zero-gap graphene
upon a 2D gapped graphene
barrier of hight $V_0$ in which the electron acquires a finite mass of $\Delta/v^2_{f}$,
as shown in Fig. \ref{fig.1}, where the tunable potential
barrier is formed by a bipolar junction ($p$-$n$-$p$) within a
single-layer graphene sheet with top gate voltage $V_g$
\cite{Huard}, $V_0$ and $d$ are the height and width of potential
barrier, respectively. Since graphene is a 2D zero-gap semiconductor
with the linear dispersion relation, $E= \hbar k_{f} v_{f}$,
the massless electrons are formally described by the Dirac-like hamiltonian
\cite{Katsnelson-NG}, $\hat{H_{0}}=-i\hbar v_{f}\sigma\nabla $,
where $v_f \approx 10^{6} m \cdot s^{-1}$ is the Fermi velocity,
$k_f$ is the Fermi wave vector ($\lambda=2 \pi/ k_f$ is Fermi wavelength), and $\sigma=(\sigma_{x}, \sigma_{y})
$ are the Pauli matrices. The wave functions of the plane wave components for
the incident and reflected regions are assumed to be
\begin{equation}
\Psi_{I}=
   \left(\begin{array}{c}
         1 \\
      se^{i\phi} \\
   \end{array}\right)
e^{i(k_{x}x+k_{y}y)}
  +r
 \left(\begin{array}{c}
      1 \\
      -s e^{-i \phi} \\
   \end{array}\right)
e^{i(-k_{x}x+k_{y}y)},
\end{equation}
so the corresponding wave function in the transmitted region can be expressed by
\begin{equation}
\Psi_{III}=t
 \left(\begin{array}{c}
      1 \\
      se^{i\phi} \\
   \end{array}\right)
e^{i(k_{x}x+k_{y}y)},
\end{equation}
where $s=sgn(E)$, $k_{x}=k_{f}\cos\phi$ and $k_{y}=k_{f}\sin\phi$ are the
perpendicular and parallel wave vector components outside the
barrier, $\phi$ is the incidence angle of the plane wave component under consideration.
For the general case of the gapped graphene barrier,
the Hamiltonian of massive Dirac fermions can be written down as
$\hat{H_{1}}=-i\hbar v_{f}\sigma\nabla +\Delta \sigma_z$, where $\Delta$ is equal to the half of the induced
gap in graphene spectrum and positive (negative) sign corresponds to the $K$ ($K'$) point, thus
the wave functions in the barrier region have the following form:
\begin{equation}
\Psi_{II}= \ \left(\begin{array}{c}
      \alpha \\
      s' \beta e^{i \theta } \\
   \end{array}\right)
e^{i(q_{x}x+k_{y}y)}
+ \left(\begin{array}{c}
      \alpha \\
      - s' \beta e^{-i \theta} \\
   \end{array}\right)
e^{i(-q_{x}x+k_{y}y)},
\end{equation}
where $s'=sgn(E-V_{0})$, $k'_f=\sqrt{(V_{0}-E)^2-\Delta^2}/\hbar v_{f}$,
$q_{x}=(k^{'2}_{f}-k_{y}^{2})^{1/2}$, $\theta=\arctan(k_{y}/q_{x})$, $\alpha$ and $\beta$ are defined by
$$
\alpha= \sqrt{1+\frac{s' \Delta}{\sqrt{\Delta^2 + k'^{2}_f}}}, ~~\mbox{and}~~
\beta= \sqrt{1-\frac{s' \Delta}{\sqrt{\Delta^2 + k'^{2}_f}}}.
$$
Accordingly, the critical angle $\phi_c$ for total reflection can be defined by
\beq
\label{critical angle}
\phi_c = \arcsin{\left[{\frac{(V_0-E)^2-\Delta^2}{E^2}}\right]^{1/2}},
\eeq
so that when $\phi > \phi_c$, the wave function in the propagating case becomes evanescent wave by replacing $q_{x}$ with
$i \kappa$, where $\kappa=(k_{y}^{2}-k^{'2}_{f})^{1/2}$.
According to the boundary
conditions, the transmission coefficient $t \equiv e^{i \varphi}/ f$ is determined by
\begin{equation}
\label{transmission}
t= \frac{1}{\cos({q_{x}d})- i (ss' \chi \sec \phi\sec \theta + \tan \phi \tan \theta)\sin(q_{x}d)},
\end{equation}
where $\chi= \sqrt{\Delta^2 + k'^{2}_f}/k'_f$ and the phase shift $\varphi$ is obtained by
\begin{equation}
\label{phase shift}
\varphi = \arctan\left[\frac{\sin\theta \sin \phi + s s' \chi}{\cos \theta \cos \phi} \tan{(q_x d)}\right].
\end{equation}
From the above expression, it is clear that in the limit $\Delta \rightarrow 0$, we get $\chi=1$ and thus one can obtain the same expression for electronic transmission probability $T= 1/f^2$ corresponding to the massless Dirac fermions \cite{XChen-APL}.

\begin{center}
\begin{figure}[]
\scalebox{0.4}[0.42]{\includegraphics{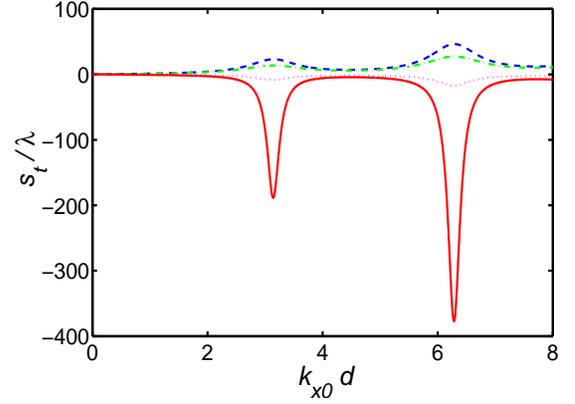}}
\caption{(Color
online) Dependence of GHL shifts in the propagating case on the barrier's width $d$, where $\phi_0= 25^\circ$, $V_{0}=120 meV$, $d$ is re-scaled to $k_{x0}d$,
$E= 80 meV$, $\Delta= 20 meV$  (solid line), $E= 220 meV$, $\Delta= 20 meV$  (dashed line),
$E= 80 meV$, $\Delta= 0 meV$  (dotted line), and $E= 220 meV$, $\Delta= 0 meV$  (dot-dashed line).
\label{fig.2}}
\end{figure}
\end{center}

\begin{center}
\begin{figure}[]
\scalebox{0.4}[0.42]{\includegraphics{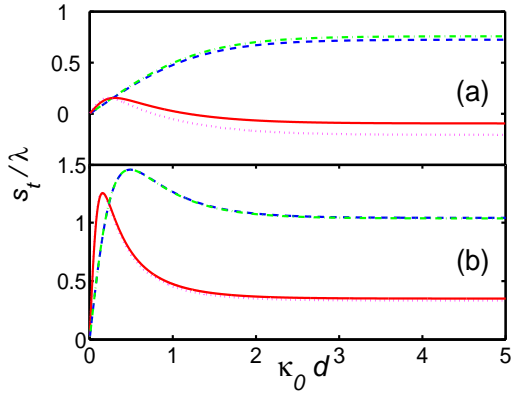}}
\caption{(Color
online) Dependence of GHL shifts in the evanescent case on the barrier's width $d$, where (a) $\phi_0= 35^\circ$ and (b) $\phi_0= 75^\circ$, $d$ is re-scaled to $\kappa_0 d$,
$E= 80 meV$, $\Delta= 20 meV$  (solid line), $E= 220 meV$, $\Delta= 20 meV$  (dashed line),
$E= 80 meV$, $\Delta= 0 meV$  (dotted line), and $E= 220 meV$, $\Delta= 0 meV$  (dot-dashed line).
\label{fig.3}}
\end{figure}
\end{center}

For a well-collimated beam with the central angle $\phi_0$ of incidence, the GHL shift can be defined, according to the stationary phase method \cite{XChen-PRB,Zhao}, as
\begin{equation}
\label{lateral shift}
s_{t} = - \frac{\partial \varphi}{\partial k_{y0}},
\end{equation}
where the subscript $0$ in this paper denotes the values taken at $\phi=\phi_0$.
The most intriguing behavior in propagating case is found for Klein tunneling, $E<V_0$,
where the GHL shifts can be negative, and also be enhanced by the transmission resonances,
whereas the lateral shifts for classical motion, $E> V_0 $, are
always large and positive.
Examples of their dependence on the barrier's width at different induced gaps $\Delta$ are
plotted in Fig. \ref{fig.2},
where $E= 80meV$, $V_{0}=120 meV$, $\phi= 25^\circ$ is less than the critical angle defined by Eq. (\ref{critical angle}), solid and dotted lines correspond to Klein tunneling $E= 80 meV < V_0$, and dashed and dot-dashed ones correspond to classical motion $E= 220 meV > V_0$. On the contrary, when the incident angle $\phi_0$ is larger than
the critical angle $\phi_c$, the lateral shifts become in the order of Fermi wavelength due to the evanescent wave, which is similar to those in total reflection at a single graphene interface \cite{Beenakker-PRL,Zhao}.
Instead of the enhancement by the transmission resonances shown in Fig. \ref{fig.2}, Fig. \ref{fig.3} illustrates that the GHL shifts for Klein tunneling and classical motion saturate respectively to negative and positive constants with increasing the barrier's width in the evanescent case, where (a) $\phi_0= 35^\circ$ and (b) $75^\circ$, $d$ is re-scaled to $\kappa_0 d$, the other parameters are the same as in Fig. \ref{fig.2}.

\section{Discussions}
In this section, we will shed light on the properties of GHL shifts in details.
For simplicity, we will find the following analytical solutions in the limit
$\Delta=0$ and discuss the negative and positive lateral shifts in cases of Klein tunneling and
classical motion, respectively.

\textit{Case 1}: Klein tunneling ($ss'=-1$). In this case, the critical angle (\ref{critical angle}) becomes
\begin{equation}
\label{critical angle-1}
\phi'_{c}=\arcsin\left(\frac{V_{0}}{E}-1\right),
\end{equation}
when the condition $E<V_{0}<2E$ is satisfied.
When the incidence angle $\phi_0$ is less then the critical angle $\phi'_{c}$ obtained above,
$\phi_{0} < \phi'_{c}$,
the lateral shift is given by
\begin{eqnarray*}
\label{displacement1}
s_t= \frac{d \tan{\phi_0}}{f^2_0}  \left\{\left[
2+ \left(\frac{k^2_{0}}{k^{2}_{x0}}+\frac{k^2_{0}}{q^{2}_{x0}}\right)\right]
\frac{\sin(2q_{x0}d)}{2q_{x0}d}-\frac{k^2_{0}}{q^2_{x0}} \right\},
\end{eqnarray*}
where $k_0=(k_{f}k'_{f}+k^2_{y0})^{1/2}$, and transmission probability $T$ is given by
Eq. (\ref{transmission}),
\begin{equation}
T \equiv \frac{1}{f^2_0} =\left[\cos^{2}(q_{x0}d)+\frac{k^{4}_{0}}
{k^{2}_{x0}q^{2}_{x0}}\sin^{2}(q_{x0}d)\right]^{-1}.
\end{equation}
The GHL shifts obtained above can be positive as well as negative, depending on the
influence of $\sin(2q_{x0}d)/(2q_{x0}d)$. Since the inequality
\begin{eqnarray}
\left[2+ \left(\frac{k^2_{0}}{k^{2}_{x0}}+\frac{k^2_{0}}{q^{2}_{x0}}\right)
\right] > \frac{k^2_{0}}{q^2_{x0}},
\end{eqnarray}
the lateral shifts can be positive only when $\sin(2q_{x0}d)/(2q_{x0}d) \rightarrow 1$ for
a very thin barrier, namely $d \rightarrow 0$. However, the lateral shifts turn negative with increasing $d$.
It is interesting that the negative lateral shifts can be enhanced by the transmission resonances.
Since the transmission coefficient is an oscillating function of tunneling parameters and
can be exhibit any value from $0$ to $1$ \cite{Katsnelson-NG}, there are resonance conditions $q_{x0}d =N \pi$, $N=0,\pm 1,\pm 2,...$ at which the barrier is transparent, $T=1$. At resonances, the lateral shifts in this case
reach $s|_{q_{x0}d =N \pi}= - (k^{2}_{0}/{q^{2}_{x0}}) d \tan{\phi_0}$, which correspond to the maximum absolute values. At anti-resonance, $q_{x0}d = (N+1/2) \pi$, the lateral shifts becomes $s|_{q_{x0}d = (N+1/2) \pi}= - ({k^{2}_{x0}}/k^{2}_{0}) d \tan{\phi_0}$. The exotic behaviors of negative and positive GHL shifts are analogous to those of lateral shifts for the transmitted light beam though left-handed metamaterial slab \cite{XChen-PRE}, based on the link between Klein paradox and negative refraction \cite{Guney}.

On the contrary, when the incidence angle is larger than the critical angle $\phi'_{c}$,
$\phi_{0} > \phi'_{c}$, the lateral shift becomes
\begin{eqnarray*}
\label{displacement2}
s_t= \frac{d \tan{\phi_0}}{f^2_0}  \left\{\left[
2+ \left(\frac{k^2_{0}}{\kappa^{2}_{0}}-\frac{k^2_{0}}{k^{2}_{x0}}\right)\right]
\frac{\sinh(2 \kappa_{0} d)}{2 \kappa_{0} d}+\frac{k^2_{0}}{\kappa^2_{0}} \right\}.
\end{eqnarray*}
In the limit of opaque barrier, $\kappa_0 d \rightarrow \infty$, the lateral shift trends to
a constant as follows,
\begin{eqnarray}
\label{displacement in the limit}
s_{t} = \frac{k_{y0}}{k_{x0} \kappa_{0}} \frac{2 k^2_{x0} \kappa^{2}_{0}-k^2_{0}(k^2_{x0}-\kappa^{2}_{0})}{k^2_{x0} \kappa^{2}_{0}+k^2_{0}},
\end{eqnarray}
which is proportional to $1/\kappa_{0}$, and implies that the GHL shift in the evanescent case is in the same order of electron wavelength as the GH effect in a single graphene interface \cite{Zhao,Beenakker-PRL}.
More interestingly, the saturated GHL shift is negative when the incidence angle satisfies
$\phi'^{c}<\phi_{0}<\phi^{\ast}$, where the critical angle is defined by
\begin{equation}
\label{critical angle-2}
\phi^{\ast}= \arcsin \sqrt{\sin \phi'_{c}}.
\end{equation}
But the GHL shift in this case will becomes positive when $\phi_{0}>\phi^{\ast}$. The
sign change of GHL shifts described by Fig. \ref{fig.3} (b) appears at the incidence angle
$\phi_{0} = \phi^{\ast}$, which is similar to the result of the quantum GH effect in graphene, taking the
the pseudospin degree into account \cite{Beenakker-PRL}.

\textit{Case 2}: classical motion ($ss'=1$). In this case, the critical angle is
\begin{equation}
\label{critical angle-2}
\phi''_{c}=\arcsin\left(1-\frac{V_{0}}{E}\right).
\end{equation}
When the incidence angle is less than the critical angle for total reflection,
$\phi_0<\phi''_{c}$, the lateral shift can be written as
\begin{eqnarray*}
\label{displacement3}
s_t= \frac{d \tan{\phi_0}}{f^2_0}  \left\{\frac{k'^2_{0}}{q^2_{x0}} + \left[
2- \left(\frac{k'^2_{0}}{k^{2}_{x0}}+\frac{k'^2_{0}}{q^{2}_{x0}}\right)\right]
\frac{\sin(2q_{x0}d)}{2q_{x0}d}\right\},
\end{eqnarray*}
where $k'_0=(k_{f}k'_{f}-k^2_{y0})^{1/2}$, and transmission probability is
\begin{equation}
T \equiv \frac{1}{f^2_0} =\left[\cos^{2}(q_{x0}d)+\frac{k^{'4}_{0}}
{k^{2}_{x0}q^{2}_{x0}}\sin^{2}(q_{x0}d)\right]^{-1},
\end{equation}
Similarly, the lateral shifts for classical motion also depend periodically on the barrier's width, thus
can be enhanced by the transmission resonances. The lateral shifts at resonances
reach $s|_{q_{x0}d = N \pi}= (k^{'2}_{0}/{q^{2}_{x0}}) d \tan{\phi_0}$, while
at anti-resonance they become $s|_{q_{x0}d = (N+1/2) \pi} = ({k^{2}_{x0}}/k^{'2}_{0}) d \tan{\phi_0}$.
However, these GHL shifts in classical motion are always positive as those in
the two-dimensional semiconductor barrier \cite{XChen-PRB}.
When $\phi_0>\phi''_{c}$, the GHL shift in the evanescent case becomes
\begin{eqnarray*}
\label{displacement4}
s_t= \frac{d \tan{\phi_0}}{f^2_0}  \left\{\frac{k'^2_{0}}{\kappa^2_{0}} + \left[
2 - \left(\frac{k'^2_{0}}{\kappa^{2}_{0}}+\frac{k'^2_{0}}{k^{2}_{x0}}\right)\right]
\frac{\sinh(2 \kappa_{0} d)}{2 \kappa_{0} d} \right\}.
\end{eqnarray*}
Then the lateral shift in the limit, $\kappa_0 d \rightarrow \infty$, is given by
\begin{eqnarray}
\label{displacement in the limit2}
s_{t} = \frac{k_{y0}}{k_{x0} \kappa_{0}} \frac{2 k^2_{x0} \kappa^{2}_{0}+k'^2_{0}(k^2_{x0}-\kappa^{2}_{0})}{k^2_{x0} \kappa^{2}_{0}+k'^2_{0}},
\end{eqnarray}
which is always positive constant.
\begin{center}
\begin{figure}[]
\scalebox{0.4}[0.42]{\includegraphics{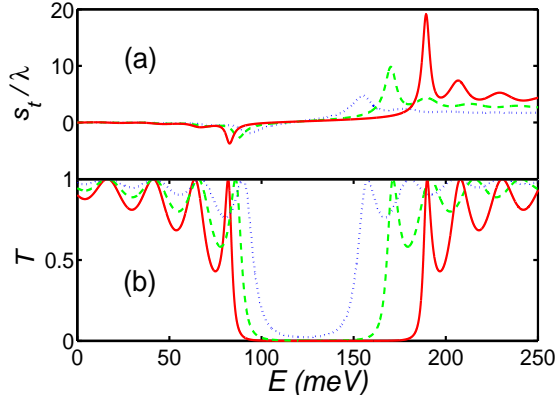}} \caption{(Color
online) GHL shifts (a) and transmission gap (b) as the function of the incident
energy E, where $d=80 nm$, $V_{0}=120 meV$, and $\Delta=0 meV$. Solid, dashed, and dotted lines
correspond to $\phi_0=20^\circ$, $15^\circ$, and $10^\circ$,
respectively.\label{fig.4}}
\end{figure}
\end{center}

Based on the properties in two cases of Klein tunneling and
classical motion, the GHL shifts (a) and corresponding transmission probabilities (b)
as the function of incidence
energy $E$ are shown in Fig. \ref{fig.4},where $d=80 nm$, $V_{0}=120 meV$, and $\Delta=0 meV$.
Solid, dashed, and dotted lines correspond to $\phi_0=20^\circ$, $15^\circ$, and $10^\circ$,
respectively. It is shown that the GHL shifts is closely related to
the transmission gap  $\Delta E =2\hbar k_y v_f$ \cite{XChen-APL}.
Fig. \ref{fig.4} indicates that the lateral shifts change the sign
near the Dirac point $E=V_{0}$, and can also be enhanced by the transmission resonances
near the boundaries of energy gap. In addition, the incidence angle has also great impact
on the GHL shifts. The absolute values of the lateral shifts increase with increasing
the incidence angles, and the positions of the maximum (absolute) values for the positive (negative) beam shifts
can also be tuned
because of the resonance conditions, as shown in Fig. \ref{fig.4}.

\begin{center}
\begin{figure}[ht]
\scalebox{0.4}[0.4]{\includegraphics{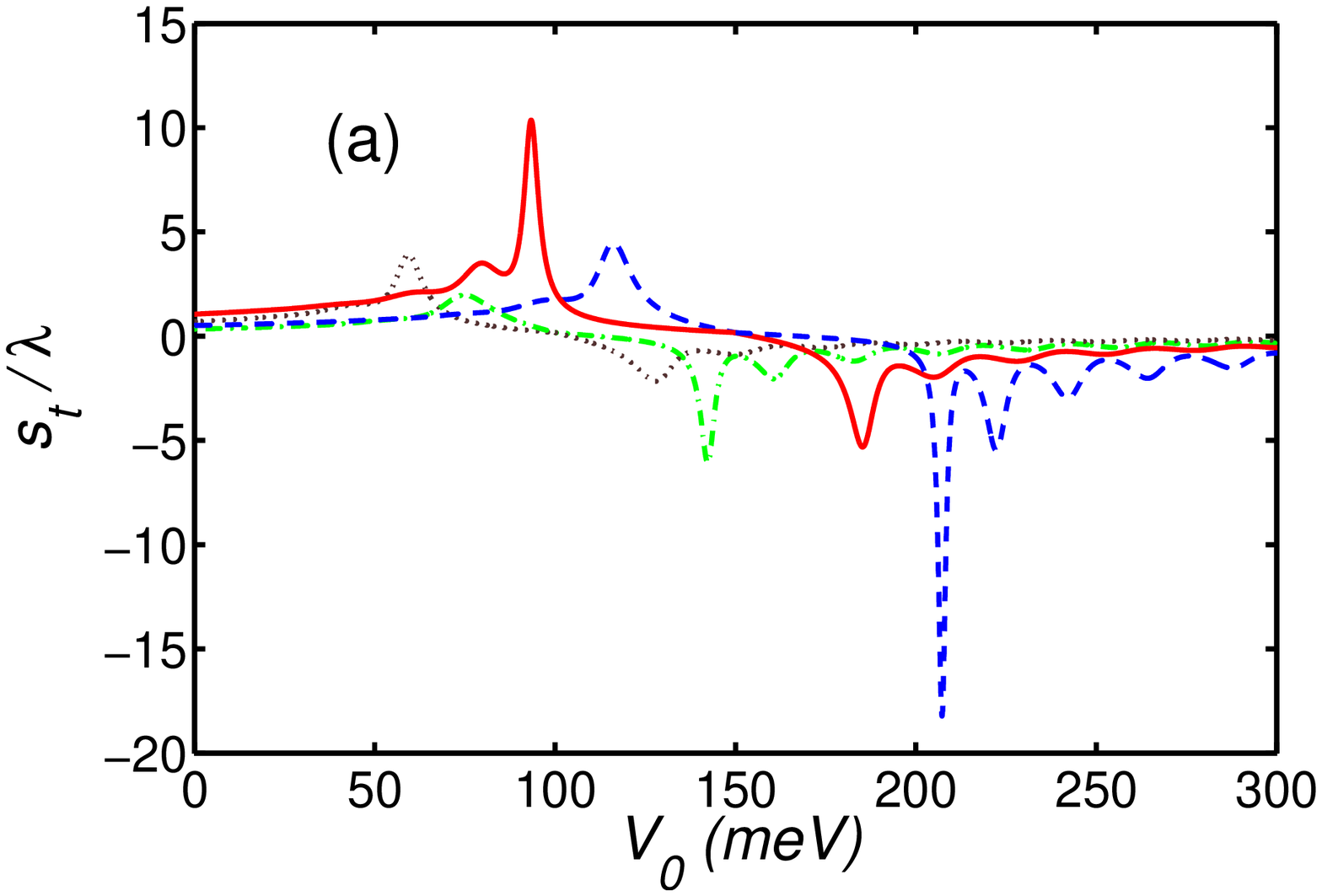}}
\scalebox{0.4}[0.4]{\includegraphics{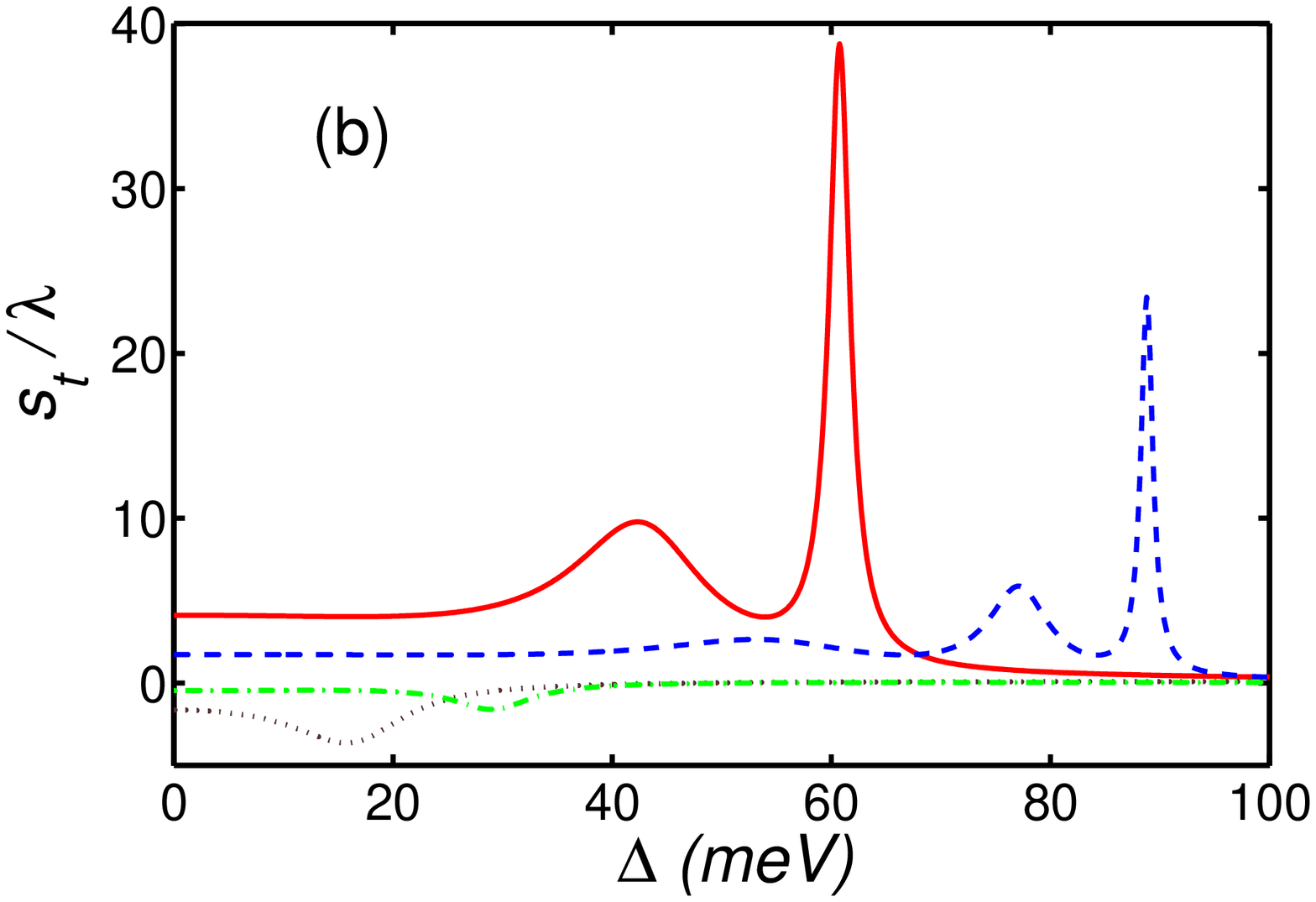}}
\caption{(Color
online) Dependence of GHL shifts on height $V_0$ of potential barrier (a) and induced gap $\Delta$ (b), where $d=80 nm$, (a) $\Delta=0 meV$, $E=150 meV$, $\phi_0=20^\circ$ (solid line), $E=150 meV$, $\phi_0=10^\circ$ (dashed line), $E=100 meV$, $\phi_0=20^\circ$ (dotted line), $E=100 meV$, $\phi_0=10^\circ$ (dot-dashed line); (b)
$V_0=120 meV$, $E=220 meV$, $\phi_0=20^\circ$ (solid line), $E=220 meV$, $\phi_0=10^\circ$ (dashed line),
$E=80 meV$, $\phi_0=20^\circ$ (dotted line), $E=80 meV$, $\phi_0=10^\circ$ (dot-dashed line).
\label{fig.5}}
\end{figure}
\end{center}

Finally, we will turn to discuss the modulations of GHL shifts by the height $V_0$ of potential barrier
and the induced gap  $\Delta$. It is shown in Fig. \ref{fig.5} that the GHL shifts can be controlled by changing the height $V_0$ of potential barrier, which can be easily implemented by applying a local top  gate voltage $V_g$ to graphene \cite{Huard}. Since the lateral shifts are in the forward and backward directions in the cases of $E>V_0$ and $E<V_0$, respectively, it is suggested that the incidence energy can be selected by these tunable beam shifts. Thus, this phenomenon does result in an alternative way to realize the graphene-based electronic devices, for example, energy splitter and energy filter. Moreover, Fig. \ref{fig.5} (b) further investigate how the GHL shifts are affected by a gap opening at the Dirac points. Comparisons of Figs. \ref{fig.2} and Figs. \ref{fig.3} further show that the energy gap will increase (decrease) the absolute values of the shifts in the propagating (evanescent) case. And the influence of gap in the propagating case is more pronounced than that in the evanescent one. The method to generate
the energy gap in graphene is through an inversion symmetry breaking of the sublattice due to the fact the densities of the particles associated with the on-site energy for A and B sublattice are different \cite{Zhou}. Therefore, the periodical dependence of GHL shifts on the gap provides an efficient way to modulate the lateral shifts in a fixed graphene barrier.

\section{Conclusion}
In summary, we have investigated the GHL shifts for Dirac fermions in transmission through a
monolayer graphene barrier. The lateral shifts, as the functions of the barrier's width
and the incidence angle, can be negative and positive in Klein tunneling and classical motion, respectively.
Since the lateral shifts have a close relation with the transmission probability, the lateral shifts
can be enhanced by the transmission resonances when the incidence angle is less than the critical angle for total reflection, while their magnitudes are only the order of Fermi wavelength
when the incidence angle is larger than the critical angle.
Compared with the smallness of conventional GH shift in graphene, the large negative and positive GHL shifts, which
can also be modulated by the height of potential barrier and the induced gap, will have potential applications in various graphene-based electronic devices. We further hope that these similar phenomena in magnetic graphene barrier may lead to the graphene-based spintronic devices on spin filter and spin beam splitter \cite{XChen-PRB,Sharma}.

\section*{Acknowledgments}
This work is supported by the National Natural Science Foundation of
China (Grants No. 60806041), the Shanghai Rising-Star Program
(Grants No. 08QA14030), the Science and Technology Commission of Shanghai
Municipal (Grants No. 08JC14097), the Shanghai Educational Development
Foundation (Grants No. 2007CG52), and the Shanghai Leading Academic Discipline
Program (Grants No. S30105). X. C. acknowledges Juan de la Cierva
Programme of Spanish MICINN and FIS2009-12773-C02-01.


\end{document}